\documentclass[12pt]{article}
\catcode`@=12
\begin{document}
\thispagestyle{empty}
\begin{flushright}
September 2002\
\end{flushright}
\vspace{0.5in}
\begin{center}
{\LARGE \bf Neutrino Radar \\}
\vspace{1.2in}

{\bf Prasanta K. Panigrahi${}^{1,2}$ and Utpal Sarkar${}^{1}$ \\}
\vspace{0.2in}

{\sl ${}^1$Physical Research Laboratory, Ahmedabad 380 009, India \\
${}^2$School of Physics, University of Hyderabad, Hyderabad, 500046, India }
\vspace{1.2in}
\end{center}
\begin{abstract}\

We point out that with improving our present knowledge of experimental
neutrino physics it will be possible to locate nuclear powered vehicles
like submarines, aircraft carriers and UFOs and detect nuclear testing. 
Since neutrinos cannot be shielded, it will not be possible to escape
these detection. In these detectors it will also be possible to perform 
neutrino oscillation experiments during any nuclear testing. 

\end{abstract}
\vspace{0.5in}

\newpage
\baselineskip 18pt

The difficulty of detecting neutrinos did not allow us to understand 
the properties of neutrinos for a long time. Only recently the 
atmospheric neutrino data has established neutrino oscillations in 
the atmospheric neutrino \cite{atm}. 
Recent results from SNO \cite{SNO}, alongwith the
other solar neutrino experiments \cite{sol}, 
have then established neutrino
oscillations in solar neutrinos. The neutrinoless double beta decay
has also been observed \cite{ndb}. 
All these results have narrowed down the
possible neutrino mass matrices for a 3-generation scenario
\cite{3gen}.

The mass squared differences between different generations of 
neutrinos are very small and it is now realised that detecting
reactor neutrinos at a distance in long baseline experiments
can probe these small mass squared differences. This interest
have now enriched the detector technology to detect neutrinos
from reactors at a large distance. The CHOOZ 
\cite{chooz} and palo Verde \cite{pv} 
reactor neutrinos have been detected at a distance of $\sim 1$ km.
In the long-baseline experiments and Kamland neutrinos will be
detected at a distance of about 200-300 kms. Detecting neutrinos from 
reactors at a distance of about 200 kms, Kamland will soon
tell us if LMA is teh solution for solar neutrino problem.

In this article we point out that with some more 
improvement of our present knowledge of
the neutrino detectors, it will be possible to think of some practical 
applications of neutrino detectors as neutrino radar. 
All the technology being developed could be used
to develop neutrino radars, which will have future defense applications.
So, developing new neutrino detectors will not only serve the
purpose of precision measurements with neutrinos, but they will
help us construct better defense radars. 

The main idea behind the neutrino radars is that all nuclear
reactors emit neutrinos in all directions and they cannot be
shielded. With proper detectors these neutrinos
could be detected at a distance. Another observation
is that the water Cherenkov detectors can also tell us the 
direction of the neutrinos. With improved 
techniques of detectors and numerical analysis like pulse
shape discrimination or the wavelet techniques, it will then be
possible to reconstruct signals in the 
detectors, which will be able to give
us real time position of the reactor source upto a certain 
distance, which depends on the strength of the source. The
source could be nuclear submarines or any nuclear powered 
war vessels or some nuclear powered
UFOs, which are not detectable otherwise. Since the range
of these detectors could be as high as few hundred to 
thousand kilometers depending on their size, they can 
definitely detect a UFO going through our atmosphere.

Another application of these detectors would be that 
if there is any nuclear testing at a large distance,
these detectors could give the exact location of the testing
site as well as the strength of the explosion. If there 
are few detectors, then it may be possible to 
do precision neutrino experiments with them during a
nuclear testing. If the testing is along the line connecting
two detectors, then the first detector can give the amount 
of neutrino flux for calibration, while the second detector
can tell us what fraction of these neutrinos have travelled
the distance between the two detectors. 
Since the neutrino flux during a nuclear
explosion will be orders of magnitude higher, these 
experiments could give us statistics, which may take several
years by reactor experiments. 

For an order of magnitude, let us first consider the range
of present detectors for detecting a nuclear vessel. 
Different detectors are planned with different aims and
none of them are aimed towards any defense applications.
So, although these detectors cannot be used as neutrino
radars, they can help us decide which will be the best
choice for the new class of detectors. 
Recently the effect of nuclear submarines on the existing
detectors have been studied \cite{sub}, 
which will be taken here as the reference point for this study. 

Nuclear reactors are used in submarines or aircraft carriers 
or other large military vessels. Thermal power of these 
vessels range between $0.3 - 1.0$ GW${}_{th}$. Exact details
for these vessels are not available, but this approximate
range is enough for the present estimate. With a 1 kton
detector mass at Kamland, a nuclear submarine at a distance
of about 40 km can produce a flux of about $10^5 
~cm^{-1}s^{-1}$ neutrinos giving about 100 counts per year, or
a aircraft carrier at a distance of 200 km could give a
flux of $10^4 ~cm^{-1}s^{-1}$ neutrinos and about 10
counts per year. In Borexino, the detector mass is about
$0.3$ kton and the detection rate will be almost one order
of magnitude less than Kamland. 

With this count rate it will not be possible to locate
any nuclear vessel with any real time analysis. So, the
present detectors may not be used for the purpose. 
However, if a few detectors are placed along the coastline,
and they are calibrated suitably, then it may be possible
to locate any nuclear vessels. Since these detections will
have directional properties, two of the detectors could
give us exact location of any vessel. If there is a third
detector, that can confirm this location and make the 
analysis easier. Coincidence between the three signals 
can be used to discriminate any background and increase
the efficiency. Since these detectors are supposed to
look for signals mostly from the sea direction, their
shape could also be different from the conventional
neutrino detectors.

The range of these detectors could be maximised if
these detectors could be placed under water at a 
distance from the coast. If the detector range is 
about 500 km, then placing the detector at a distance
of 500 km from the coast would cover a range of about
1000 km. This warrants developments of underwater 
neutrino detectors. 

Since the range of these detectors are expected to be 
few hundred kms, any nuclear powered aircraft will be 
within the range of these detectors when they enter
our atmosphere in the vicinity of the detectors. So,
if any UFOs fly over the earth at a height beyond the 
reach of any ordinary radars, they may escape
detection by the radars, but may get detected by
the neutrino radars. 

Taking a crude estimate of the power generated in a nuclear
testing to be $4-6$ orders of magnitude higher than 
the power generated in a nuclear reactor, if any
nuclear testing takes place at a distance of about 
200 km from the detector, then there will be an 
increase in the neutrino count by the same orders of
magnitude. Thus a testing taking place at a distance
of about 20,000 km can also be detected by these 
detectors. 

If a nuclear testing takes place at a distance of
say, 1000 km, then the neutrino flux at these
detectors will be at least two orders of magnitude
higher than the flux expected from any reactors at
a distance of a few km. So, if there are three detectors
placed along the line of the testing site, then it may
be possible to calibrate the flux by one of the 
detectors and then perform neutrino oscillation 
experiments with the data available from the next
two detectors. The statistics will be much
higher than any reactor neutrino experiments, which is
important for any precision measurements.

As a concrete example, we consider $\bar{\nu_{e}}$ flux from localized 
sources such as reactors powering the nuclear vehicles
or the same originating from nuclear blasts, as a potential candidate 
for detection. The reaction 
$\bar{\nu_{e}}+P~ \rightarrow~e^{+}~+~N$ can, not only be 
utilized in large liquid scintillation detectors, but also gives 
a positron coincidence tag at $\sim0.2$ ms delay, which can be used 
for suppressing the background.   
Other major background inteferences, originating from known reactors, muon decays in 
the atmosphere, supernovae relics and 
geo-${\bar \nu_{e}}$'s from the decay of $^{238}\bf{U}$ and $^{232}\bf{Th}$ 
\cite{ragh} can, in principle, be  
discriminated, firstly because of the assumed transient nature of the fluxes being 
looked for and also due to the spectral separation of the neutrinos of 
reactor and other origins. For example, the  geo-${\bar \nu_{e}}$ events have a 
signal window from $1.02-3.26$ MeV, whereas the reactor neutrinos extend 
much beyond that. In fact, the precisely calculable spectral shape and flux from 
known reactors can be used for calibration purposes, against which the transients 
can be identified. It should also be mentioned that, non-$\bar{\nu_{e}}$ background 
can also originate from cosmic ray and other events mimicking the tag, which should 
be separated, either through concident measurements or via pulse shape analysis.         

The major cause of reduction of fluxes from unknown compact sources is neutrino
oscillation, through which ${\bar \nu_{e}}$'s can be converted to 
${\bar \nu_{\mu,\tau}}$. Better detectors and more accurate measurements of 
${\Delta m^2={m_1}^2-{m_2}^2}$ and $\sin^2{2\theta}$ in future, 
where $m_{1,2}$ are the eigenstate masses, can convert this potential 
weakness into a strength, by observing the survival spectrum $F^C(E_{\nu})$ at different 
distances, as a function of energy $E_{\nu}$. The source distance $r$ and location 
can be estimated from $F^C(E_{\nu})=\int dr F(E_{\nu},r)[1~-\sin^2{2\theta}\sin^2[(r/4)\Delta m^2/E_{\nu}]$, 
by observation through multiple detectors. 
One can also think of detection of the neutrinos belonging to the first two flavors via 
elastic scattering, which does not distinguish between them.

The transient nature of the fluxes from moving sources, as also the different spectral shapes of 
various signals, suggests the use of wavelets \cite{da}
for the purpose of data analysis. The wavelet decomposition of the signals, in terms of 
scale and translation variables, may turn out handy, for separating different signals in the 
spectral domain. Wavelets are ideal for handling irregular data series
\cite{gen}. Given the observed signal, 
as a function of time, $y(t)=f(t)+e(t)$, where $f(t)$ is the signal and $e(t)$ is the noise, Donoho and 
co-workers have shown \cite{don} the usefulness of the wavelets, in extracting $f(t)$, when the noise is below 
certain threshold and the signal variation is well above it. This is achieved through appropriate 
threshholding in the domain of wavelet coefficients. Since, the presence of transient signals
produces significant signal variations, wavelets may find profitable application in the analysis of 
neutrino signals. It should be emphasized that, traditional methods like local smoothing, for extracting 
the signal will not work for the cases where, the signal is quite irregular.

In summary, we point out that improvement of the present 
technology of neutrino detectors may allow us to use
these detectors for defense purposes. The new neutrino radars 
will then be capable of detecting 
{\it Nuclear powered submarines};
{\it Nuclear powered aircraft carriers or any other
army vessels}; 
{\it Nuclear powered UFOs}; 
{\it Nuclear testing (with the information of the 
strength of the explosion)};
They may also be used to do neutrino oscillation experiments
during any nuclear testing. Although these applications are
not possible with the available and planned detectors, but
all the available knowledge on neutrino detectors will be
needed to design these new class of detectors. In this 
article we do not present any details about the type of 
these detectors, rather we emphasize that our present
knowledge of neutrino detectors should be enriched by
constructing more detectors, considering
the possibility of all these defense applications. 
So, even if the solar neutrino problem is settled soon,
we cannot afford to stop experimental neutrino physics in
the near future. 

\bibliographystyle{unsrt}

\begin{thebibliography}{99}
\bibitem{atm} S. Fukuda {\it et al.}, Super-Kamiokande Collaboration,
Phys. Rev. Lett. {\bf 85}, 3999 (2000) and references therein.
\bibitem{SNO} Q.~R.~Ahmad {\it et al.}, SNO Collaboration, Phys. Rev. Lett.
{\bf 89}, 011301, 011302 (2002).
\bibitem{sol} S.~Fukuda {\it et al.}, Super-Kamiokande Collaboration,
Phys. Rev. Lett. {\bf 86}, 5656 (2001) and references therein; Q.~R.~Ahmad
{\it et al.}, SNO Collaboration, Phys. Rev. Lett. {\bf 87}, 071301 (2001).
\bibitem{ndb} H. V. Klapdor-Kleingrothaus {\it et al.}, Mod. Phys. Lett. 
{\bf A16}, 2409 (2001).  
\bibitem{3gen} F. Vissani, hep-ph/9708483; R. Adhikari and
G. Rajasekaran, Phys. Rev. {\bf D 61} (1999) 031301(R);
H. Georgi and S.L. Glashow, Phys. Rev. {\bf D 61} (2000) 097301;
 H.V. Klapdor-Kleingrothaus and U. Sarkar, Mod. Phys. Lett. {\bf A 16}
(2001) 2469; Phys. Lett. {\bf B 532}
(2002) 71; V. Barger, S.L. Glashow, D. Marfatia and
  K. Whisnant, Phys. Lett. {\bf B 532} (2002) 15;
V. Barger, S.L. Glashow, P. Langacker and D. Marfatia, Phys. Lett.
{\bf B 540} (2002) 247; Y. Uehara, Phys. Lett. {\bf B 537} (2002{\bf B 540} (2002) 247; Y. Uehara, Phys. Lett. {\bf B 537} (2002) 256;
F. Feruglio, A. Strumia and F. Vissani, Nucl. Phys. {\bf B637} (2002) 345;
E. Ma and G. Rajasekaran, Phys. Rev. {\bf D64} (2001) 113012;
E. Ma, Mod. Phys. Lett. {\bf A17} (2002) 289;
Mod. Phys. Lett. {\bf A17} (2002) 627;
K. S. Babu, E. Ma, and J. W. F. Valle, hep-ph/0206292;
Z. Xing, Phys. Rev. {\bf D 65} (2002) 077302.
\bibitem{chooz} M. Apollonio {\it et. al.}, Phys. Lett. {\bf B 466}, 415 (1999).
\bibitem{pv} F. Boehm {\it et. al.}, Phys. Rev.
{\bf D 64}, 112001 (2001).
\bibitem{sub} J. Detwiler, G. Gratta, N. Tolich and Y. Uchida,
hep-ex/0207001.
\bibitem{bbmass} See for example H. V. Klapdor-Kleingrothaus and U. Sarkar,
Phys. Lett. {\bf B532}, 71 (2002); S. Pascoli and S. T. Petcov,
hep-ph/0205022; and references therein.
\bibitem{a4} E. Ma and G. Rajasekaran, Phys. Rev. {\bf D64}, 113012 (2001);
E. Ma, Mod. Phys. Lett. {\bf A17}, 289 (2002); E. Ma, Mod. Phys. Lett.
{\bf A17}, 627 (2002).
\bibitem{ragh} R. Raghavan {\it et. al.}, Phys. Rev. Lett. {\bf 80}
(1998) 635; L.M. Krauss, S.L. Glashow and D.N. Schramm, 
Nature {\bf 310} (1984) 191.
\bibitem{da} I. Daubechies, {\it Ten lectures on wavelets}, vol. {\bf 64}
of {\it CBMS-NSF regional conference series in Applied Mathematics, Society 
for Industrial and Applied mathematics}, Philadelphia,
1992.
\bibitem{gen} R. Gencay, F. Selcuk, B. Whitcher, {\it An Introduction
to wavelets and other filtering methods in finance and economics}, 
Academic Press, 2001.
\bibitem{don} D. Donoho, I. Johnstone, G. Kerkyacharian and 
D. Pichard, Jour. of Roy. Stat. Soc. {\bf 57} (1995) 301.
\end{thebibliography}

\end{document}